\pdfoutput=1

\documentclass[11pt]{article}

\usepackage[]{ACL2023}

\usepackage{times}
\usepackage{latexsym}

\usepackage[T1]{fontenc}

\usepackage[utf8]{inputenc}

\usepackage{microtype}

\usepackage{amsmath}
\usepackage{tabularx}
\usepackage{booktabs}
\usepackage{multirow}
\usepackage{array}
\usepackage{colortbl}
\usepackage{xspace}

\usepackage{inconsolata}
\usepackage{graphicx} 
\graphicspath{ {figures/} }

\newcommand{\scincl}{\textsc{SciNCL}\xspace}
\newcommand{\qpcr}{\textsc{qPCR}\xspace}

\title{Paragraph-level Citation Recommendation \\based on Topic Sentences as Queries}

\author{Zoran Medi\'{c} \and Jan \v{S}najder\\
University of Zagreb, Faculty of Electrical Engineering and Computing\\
Text Analysis and Knowledge Engineering Lab\\
\tt \{zoran.medic,jan.snajder\}@fer.hr
}

\begin{document}

\maketitle

\begin{abstract}
Citation recommendation (CR) models may help authors find relevant articles at various stages of the paper writing process.
Most research has dealt with either global CR, which produces general recommendations suitable for the initial writing stage, or local CR, which produces specific recommendations more fitting for the final writing stages.
We propose the task of \emph{paragraph-level CR} as a middle ground between the two approaches, where the paragraph's topic sentence is taken as input and recommendations for citing within the paragraph are produced at the output. 
We propose a model for this task, fine-tune it using the quadruplet loss on the dataset of ACL papers, and show improvements over the baselines. 
\end{abstract}

\section{Introduction}

When writing papers, scientists should position their work within the landscape of existing work and cite relevant articles.
However, keeping track of related work is becoming increasingly taxing in the era of information overload \citep{johnson2018stm}. Citation recommendation (CR) models may help the authors by producing a list of articles to be cited for a given query. 
In the task of \emph{global citation recommendation} (GCR), \citep{bhagavatula-etal-2018-content,ali2021global,ali2022spr} the queries are typically composed of the citing article's title and abstract. 
In contrast, in \emph{local citation recommendation} (LCR) \citep{ebesu2017neural,medic-snajder-2020-improved,gu2022local} queries are citation contexts from the citing article's body. 
Consequently, while GCR recommendations tend to be diverse and general, LCR recommendations are typically more specific and tailored to the citation context. 
These different types of recommendations respond to different information needs in the process of writing a scientific article.
GCR recommendations are perhaps more beneficial in the initial stages of research when the authors have a rough research idea and look for previous work on that topic.
In contrast, LCR recommendations are better suited for the final stages of the writing process when the authors are already familiar with prior work and only want to complement the text with citations at specific places.
However, between the initial and the final stages, there are typically many intermediate stages where authors draft and revise their manuscripts. 
Gauging the usability of GCR and LCR for these intermediate stages is difficult, where queries are likely to be more specific than for GCR but less specific than for LCR. 

In this work, we consider a CR approach targeting the intermediate stages of the writing process, particularly the writing of the related work (RW) section. 
When writing the RW section, authors typically know enough about the topic to 
identify and name the main strands of RW but cannot yet produce detailed citation contexts.
We introduce the \textbf{P}aragraph-level \textbf{C}itation \textbf{R}ecommendation (\textbf{PCR}) task, which takes a paragraph's topic sentence as a query and produces a list of articles for citing in that specific paragraph.
As topic sentences are typically more general than citation contexts appearing later in the paragraph, they can be seen as proxies for queries the author might pose in the intermediate stages of writing a RW section.

We develop the PCR model by fine-tuning \scincl \citep{ostendorff2022scincl}, a state-of-the-art transformer-based article encoder. 
We fine-tune \scincl on a dataset derived using CORWA \citep{li-etal-2022-corwa}, a dataset of sentences from the related work sections annotated with discourse labels. 
The proposed model performs better than a baseline GCR model on this task, demonstrating the potential of  paragraph-level citation recommendation. 
Our contributions are (1) PCR, a novel task of producing citation recommendations for an article's paragraph based on the paragraph's topic sentence and (2) a model for PCR fine-tuned and adapted from the existing GCR state-of-the-art model. 
\section{Related Work}
\label{sec:rw}

State-of-the-art CR models all rely on deep learning.
Typically, the models encode text separately from the metadata information such as authors, venues, or citation graphs \citep{bhagavatula-etal-2018-content,medic-snajder-2020-improved,ali2021global}.
In recent work on GCR and LCR, the text is typically encoded with transformer-based modules \citep{vaswani2017attention}, which produce article embeddings that can be used in the different recommendation and classification tasks \citep{cohan2020specter,ostendorff2022scincl}.
For an overview of the approaches to CR tasks, we refer the reader to \citep{medic2020survey}.
We follow the recent line of work that uses transformer-based encoders to produce query and candidate article embeddings.
However, our work differs from previous work in that our model produces recommendations at the paragraph level. 

Recently introduced large language models \citep{brown2020language,zhang2022opt,scao2022bloom,taylor2022galactica} that generate text when prompted could be adapted for CR tasks.
CR queries could be used to construct prompts that require the model to output a list of recommended articles or even generate an article's text with citations inserted.
However, such models tend to generate texts containing false information \citep{ji2022survey}. In this work, we focus on the task of retrieving relevant articles given a query, leaving the generation of citing text for future work.

\section{Paragraph-Level CR}
\label{sec:task}

\paragraph{Task.}

As with GCR and LCR, we define PCR as a ranking task: 
given a query $\mathit{q}$ and a pool of articles $\mathit{P} = \{\mathit{a}_{(i)}\}_{i=1}^{N}$, the model has to rank the articles in $\mathit{P}$ based on the recommendation score for citing in $\mathit{q}$. 
To construct a query $\mathit{q}$, we consider a citing article $\mathit{a_q}$, consisting of a title and an abstract, a paragraph $\mathit{p_q}$ from $\mathit{a_q}$, and sentences $[\mathit{s_1,...,s_m}]$ that make $\mathit{p_q}$.
We form $\mathit{q}$ by concatenating $\mathit{a_q}$'s title and abstract with the sentence $\mathit{s_1}$ from $\mathit{p_q}$, i.e., paragraph's \textit{topic sentence}.
In this setup, the model performs successfully if it ranks high the articles that are cited in the rest of $\mathit{p_q}$ (i.e., in $[\mathit{s_2,...,s_m}]$). 

\paragraph{Dataset.}

To identify topic sentences in paragraphs, we used CORWA \citep{li-etal-2022-corwa}, a dataset derived from S2ORC \citep{lo2020s2orc}, a large dataset of over 81M articles from various scientific fields. 
CORWA contains discourse-labeled sentences from the RW sections of ACL articles.\footnote{Association for Computational Linguistics, \url{https://aclanthology.org/}} 
Albeit citations appear in paragraphs from other sections as well, we focused on RW sections as they provide citations to work topically relevant to the citing article and typically group them into paragraphs by topic.
CORWA's discourse labels reflect the sentences' specific roles and information sources.
We rely on the \textit{Transition} label, assigned to the sentences that ``serve as topic introductions or transitions from one topic to another'' \citep{li-etal-2022-corwa}.
Suppose the first sentence in a paragraph is labeled as \textit{Transition}. 
In that case, we considered it a topic sentence and treated the citations in the rest of the paragraph as relevant for a query constructed using that sentence.
Filtering CORWA based on this criterion left us with 790 annotated RW paragraphs that start with a topic sentence and cite at least one article. 
To enlarge the dataset, we used the discourse classifier from \citep{li-etal-2022-corwa} and applied it to all the ACL articles with full text available in S2ORC.
We then added the silver-labeled paragraphs that started with a ``\textit{Transition}''-labeled sentence to the gold-labeled dataset, resulting in a total of 10,481 paragraphs.

To construct the training and evaluation sets, we split the paragraphs published before, in, and after 2017 into the train, validation, and test set, respectively.
This resulted in 7,773, 560, and 2,148 paragraphs in the train, validation, and test set, respectively. 
For the pool of articles for test set evaluation, we extracted from S2ORC all articles with an ACL ID and all the articles cited in any ACL article.
We also removed all the articles with empty title or abstract fields and ensured that only articles older than the query make up the pool. The resulting pool contained 94,129 articles.

\section{Model}
\label{sec:model}

We based the PCR model on \scincl \citep{ostendorff2022scincl}, a transformer-based scientific articles encoder that obtained state-of-the-art results on benchmarks for evaluating scientific article representations \citep{cohan2020specter,medic-snajder-2022-large}. 
The model takes an article's title and abstract as input and produces an article embedding. 
It has been trained using triplet loss with triplets of the form $\mathit{(a^{q}, a^{+}, a^{-})}$, where $\mathit{a^{q}}$ is the query (the citing article), $\mathit{a^{+}}$ is a positive item (a cited article), and $\mathit{a^{-}}$ is a negative item (an article not cited in the citing article).
Given embedding $\mathbf{e}$ for input $a$, the triplet loss is defined as
$\mathcal{L}_t = \max(0, \| \mathbf{e^{q}} - \mathbf{e^{+}} \| - \| \mathbf{e^{q}} - \mathbf{e^{-}} \| + m)$,
where $\|\cdot\|$  is the L2 norm and $m$ is a margin between the two norms.
We adapted \scincl to PCR by modifying the query construction and the loss function. 
We built queries by concatenating the paragraph's topic sentence to the citing article's title and abstract and separating it from the rest with a special \texttt{[TS]} token.
As in \scincl, we took the final layer's \texttt{[CLS]}'s representation as the query's embedding.
We used the same encoder to embed the pool articles, with their title and abstract as input.

Unlike \citet{ostendorff2022scincl}, who minimize the triplet loss, we minimized the quadruplet loss.
The quadruplet loss has been used for LCR \citep{zhang2021recommending}, where it has been found to outperform models trained with triplet loss.
The quadruplet has two positive items, $\mathit{a^{+}_{1}}$ and $\mathit{a^{+}_{2}}$, and the loss is modified accordingly.
Given a quadruplet $(a^{q}, a^{+}_{1}, a^{+}_{2}, a^{-})$, we minimized the following loss:
\begin{align*}
\small\label{eq:quad-loss}
\mathcal{L}_q &= \sum_{i} \max(0, \| \mathbf{e^{q}} - \mathbf{e^{+}_{i}} \| - \| \mathbf{e^{q}} - \mathbf{e^{-}} \| + m) \\ 
&+ \sum_{i} \max(0, \| \mathbf{e^{+}_{1}} - \mathbf{e^{+}_{2}} \| - \| \mathbf{e^{+}_{i}} - \mathbf{e^{-}} \| + m)
\end{align*}
Compared to $\mathcal{L}_t$, loss $\mathcal{L}_q$ includes an additional term for promoting similarity between the query and the second positive item (the first term), as well as two additional terms that force the embeddings of the positive items to be close to each other and far from the negative item (the second term).

Previous work has shown the importance of choosing informative and hard negative examples for training text encoders \citep{li2019sampling,ostendorff2022scincl}. 
To select the challenging negative items, we sampled them from three pools: (1) articles cited in the RW section of the citing article but in a paragraph different from the query's, (2) articles cited in the citing article but in a section other than RW, and (3) articles cited in the cited articles but not in the citing article. 
We sampled 10 quadruplets for each paragraph: three quadruplets each from the first and the second pool, and four quadruplets from the third pool.

We fine-tuned the \scincl checkpoint from the Hugging Face Hub \citep{wolf-etal-2020-transformers} for five epochs and tracked R-precision on the validation set.
To avoid overfitting, we only fine-tuned the last two layers of \scincl.
We used the AdamW optimizer \citep{loshchilov2018decoupled} with $\beta_1$ = 0.99, $\beta_2$ = 0.999, weight decay set to 0.01, and learning rate of 1e-5 with a linear scheduler for the first 10\% of learning steps.
The value of $m$ in $\mathcal{L}_q$ was set to 0.5 for all the loss terms.
We embedded all the test set queries and the pool articles using the trained model and performed the nearest neighbor search using Faiss \citep{johnson2019billion}.
We call our model \textbf{\qpcr} (\textbf{q}uadruplet-based \textbf{P}aragraph \textbf{C}itation \textbf{R}ecommender).\footnote{Our code with links to data and the model is available at: \url{https://github.com/zoranmedic/qpcr}}

\section{Experiments}
\label{sec:results}

We compared the performance of \qpcr against three baselines: (1) \scincl with the citing article's title and abstract as input (\textsc{SciNCL-base}\xspace), (2) \scincl with topic sentence concatenated with title and abstract as input (\textsc{SciNCL-ts}\xspace), and (3) \scincl fine-tuned with ACL's RW triplets (\textsc{SciNCL-aclrw}\xspace).
For the last model, we fine-tuned \scincl with triplets of cited and non-cited pairs sampled from the articles cited in the RW sections of ACL articles.
This baseline helps us analyze whether the performance gap between \qpcr and \textsc{SciNCL-ts}\xspace is due to domain adaptation (as \scincl was trained on a multi-domain corpus) or the inclusion of a topic sentence.
We also report the results of fine-tuning \scincl with triplets instead of quadruplets (\textsc{SciNCL-tri}\xspace).

We report R-precision, R@5, R@10, and MRR.
Among these, R-precision seems the most appropriate for the PCR task, as the number of cited articles varies across paragraphs.\footnote{The average number of articles cited per paragraph is 3.5.} 
However, since authors often do not cite all the relevant articles, but only a few, we also report R@5 and R@10 as more lenient metrics that do not penalize false positives as much as R-precision.
We report an average across three fine-tuning runs for each model we fine-tuned.

\begin{table}[t]
\small
\renewcommand{\arraystretch}{1.4}
\centering
\tabcolsep=0.28cm
\begin{tabular}{@{}lcccc@{}}
\toprule
Model & R-prec & R@5 & R@10 & MRR \\ \midrule
\textsc{SciNCL-base}\xspace & 6.92 & 8.47 & 12.70 & 17.32  \\
\textsc{SciNCL-ts}\xspace & 7.62 & 9.86 & 14.58 & 19.23  \\
\textsc{SciNCL-aclrw}\xspace & 7.61 & 9.87 & 14.77 & 19.21 \\
\textsc{SciNCL-tri}\xspace & 7.63 & 9.57 & 14.58 & 19.30 \\
\qpcr & \textbf{8.00} & \textbf{10.21} & \textbf{15.19} & \textbf{20.05} \\ 
\bottomrule
\end{tabular}
\caption{Performance of PCR models on test set queries. Values are in the 0--100 range. The best results are shown in \textbf{bold}.}
\label{tab:results}
\end{table}

\paragraph{Results.}

Table~\ref{tab:results} shows the results.
The first observation is that \textsc{SciNCL-ts}\xspace yields better results than \textsc{SciNCL-base}\xspace, demonstrating the potential of using topic sentences in CR queries.
\textsc{SciNCL-tri}\xspace does not perform better than \textsc{SciNCL-ts}\xspace, while \qpcr obtains the best results across all the reported metrics.
The increase in performance from \textsc{SciNCL-tri}\xspace to \qpcr indicates the benefits of training with a more informative signal derived from quadruplets instead of triplets.
Another observation is that domain adaptation of \scincl on ACL RW triplets does not increase performance compared to \textsc{SciNCL-ts}\xspace, except for R@10.

Overall, the improvement of \qpcr over \textsc{SciNCL-ts}\xspace, the best-performing baseline model that does not use topic sentences in training, is not that large: ranging from 0.35 for R@5 to 0.82 for MRR. 
We believe this small difference is indicative of the difficulty of the task.
It also reflects an issue with using citations as relevance signals when articles are represented using text only, as text is often not the only citing criteria \citep{mammola2022measuring}.

\paragraph{Qualitative Analysis.}

\begin{figure}
    \centering
    \includegraphics[scale=0.5]{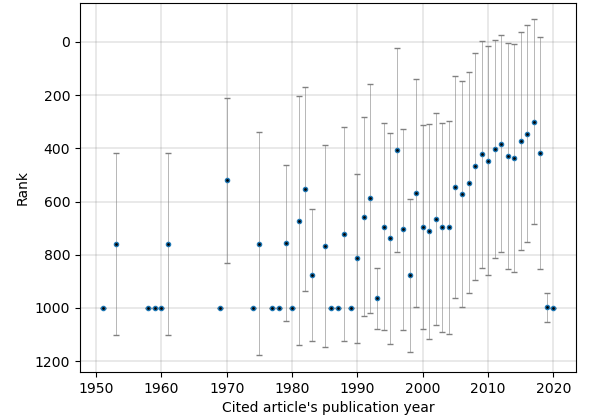}
    \caption{Average ranks of cited articles per publication year for the ranking obtained with \qpcr.}
    \label{fig:rank_years}
\end{figure}

To gain insights into PCR's difficulty and \qpcr's performance, we manually compared some of the \qpcr and \textsc{SciNCL-ts}\xspace ranking lists.
Recommendations provided by both models often seem relevant to the given paragraph but are counted as false positives as they were not cited in the paragraph (cf.~Appendix for some examples).
While this is not surprising, as citing all articles relevant to a specific context is not feasible, it suggests that better evaluation datasets are needed for developing CR models.

We also find that both models perform worse when the cited article is older than the citing article. 
Fig.~\ref{fig:rank_years} shows the average \qpcr ranking of cited articles across their publication years, revealing that older articles tend to have a lower ranking. 
In contrast, new ones are closer to the top of the ranking list.
This is also confirmed by the positive Pearson correlation coefficient of 0.15 (p$<$0.05) between the difference in cited and citing article's publication years ($\Delta t)$ and the cited article's rank. 
We hypothesized this might be due to the vocabulary mismatch between the citing and cited articles that developed over time (cf.~Appendix for some examples). 
Indeed, the Pearson correlation coefficient between $\Delta t$ and the Jaccard index between the query as defined in PCR and the cited article (as a measure of word overlap) is $-0.24$ (p$<$0.05), i.e., a larger gap in publication years is accompanied by a smaller term overlap, supporting (though not proving) the vocabulary mismatch hypothesis. 

Another observation is that \qpcr more often than \textsc{SciNCL-ts}\xspace recommends articles with titles more topically aligned with a paragraph's topic sentence (cf.~Appendix for some examples).
To verify this, we calculated the L2 distance between the embeddings of topic sentences and the titles of recommended articles using CoSentBert \citep{mysore-etal-2022-multi}, a variant of Sentence-BERT \citep{reimers-gurevych-2019-sentence} trained on the scientific domain.
The average distance between topic sentences and titles recommended by \qpcr and \textsc{SciNCL-ts}\xspace is 14.16 and 14.38, respectively.
In addition, on average, \qpcr's attention heads across all layers assign higher weights to tokens from the topic sentences than \textsc{SciNCL-ts}\xspace.
Specifically, the average sum of the topic sentence's token weights in the case of \qpcr and \textsc{SciNCL-ts}\xspace is 0.25 and 0.12, respectively.
These findings confirm that \qpcr leverages topic sentence information from the query. 
We leave the investigation of what constitutes an optimal PCR query for future work.

\section{Conclusion}
\label{sec:conclusion}

We introduced paragraph-level citation recommendation (PCR), a task of recommending articles for citing in a paragraph based on the paragraph's topic sentence. 
The proposed \qpcr model outperformed other sensible baselines and demonstrated improvements of training with the quadruplet loss over standard triplet loss. 
Interesting future work directions include generative PCR models, which use \qpcr's recommendations to generate paragraphs, and the construction of evaluation datasets based on stronger relevance signals than  citations.

\section*{Limitations}

\qpcr was trained only with paragraphs from RW sections of ACL articles.
This raises the question of how well the model would perform in different article sections and scientific domains.
Furthermore, as our dataset relies on CORWA's classifier for discourse labels, errors made by the classifier are propagated to our dataset and can affect the model's performance.
Additionally, the articles' texts are processed using parsers and tools for sentence segmentation that could result in errors and produce both incoherent titles and abstracts and topic sentences identified by CORWA's classifier.
Regarding the language, our dataset contains only the articles written in English, which limits the applicability of our model to other languages.

\section*{Ethics Statement}

Both the dataset we use and the model we fine-tune are publicly available.
Our experiments' dataset is constructed from CORWA \citep{li-etal-2022-corwa}, derived from S2ORC \citep{lo2020s2orc}, a publicly available dataset of scientific articles released under CC BY-NC 2.0 license.
Silver-labeled part of the dataset in our experiments was obtained by running CORWA classifier on articles from S2ORC.
\scincl, the model that we fine-tune and use in our experiments, is available at Hugging Face Hub \citep{wolf-etal-2020-transformers} and released under MIT Licence.

As we use citations from S2ORC as labels when training \qpcr, the bias inherent to citing patterns in articles from S2ORC might be passed onto the model trained on these articles.
For example, training the model on articles that cite the work from specific authors or institutions when more articles fit into the context might lead to the model preferring articles from particular authors or institutions over others.
Many other subjective factors influencing citing behavior might be passed onto the model trained on such data.
However, despite the potential biases that might appear in the model we trained, we believe introducing a new task in our work benefits the scientific community. 
It could stimulate more work on the task of citation recommendation and, alongside it, the work on debiasing citation recommendation systems in various scenarios.

\section*{Acknowledgments}

The first author has been supported by a grant from the Croatian Science Foundation (HRZZ-DOK-2018-09).

\bibliography{custom}
\bibliographystyle{acl_natbib}

\appendix

\section{Appendix}

Here we provide examples of paragraphs in which older articles were cited and of those for which many of the recommendations by either \textsc{SciNCL-ts}\xspace or \qpcr seem relevant but were considered as false positives since they were not cited in the given paragraph.
Examples of paragraphs with older cited articles are given in Tables \ref{tab:example-2.1} and \ref{tab:example-2.2}.
Examples of cases with seemingly relevant false positives are given in Tables \ref{tab:example-3}, \ref{tab:example-4}, and \ref{tab:example-5}.
Tables \ref{tab:example-6} and \ref{tab:example-7} showcase examples of paragraphs for which \qpcr produces recommendations with titles that are more topically relevant for the topic sentence than those of \textsc{SciNCL-ts}\xspace.

\begin{table*}
\small
\centering
\begin{tabular}{|p{0.05\textwidth}|p{0.38\textwidth}|p{0.05\textwidth}|p{0.38\textwidth}|}
\hline
\multicolumn{4}{|l|}{\textbf{Title:} CommonsenseQA: A Question Answering Challenge Targeting Commonsense Knowledge} \\
\hline
\multicolumn{4}{|l|}{\textbf{Year:} 2018} \\
\hline
\multicolumn{4}{|p{0.97\textwidth}|}{\textbf{Abstract:} When answering a question, people often draw upon their rich world knowledge in addition to some task-specific context. Recent work has focused primarily on answering questions based on some relevant document or content, and required very little general background. To investigate question answering with prior knowledge, we present CommonsenseQA: a difficult new dataset for commonsense question answering. To capture common sense beyond associations, each question discriminates between three target concepts that all share the same relationship to a single source drawn from ConceptNet (Speer et al., 2017). This constraint encourages crowd workers to author multiple-choice questions with complex semantics, in which all candidates relate to the subject in a similar way. We create 9,500 questions through this procedure and demonstrate the dataset's difficulty with a large number of strong baselines. Our best baseline, the OpenAI GPT (Radford et al., 2018), obtains 54.8\% accuracy, well below human performance, which is 95.3\%.
} \\
\hline
\multicolumn{4}{|p{0.97\textwidth}|}{\textbf{Paragraph:} \textit{Machine common sense, or the knowledge of and ability to reason about an open ended world, has long been acknowledged as a critical component for natural language understanding.} Early work sought programs that could reason about an environment in natural language \textcolor{olive}{(McCarthy, 1959)} , or leverage a world-model for deeper language understanding (Winograd, 1972) . Many commonsense representations and inference procedures have been explored (McCarthy and Hayes, 1969; Kowalski and Sergot, 1986 ) and large-scale commonsense knowledge-bases have been developed (Lenat, 1995; Speer et al., 2017) . However, evaluating the degree of common sense possessed by a machine remains difficult.}\\
\hline
Rank & \textbf{\textsc{SciNCL-ts}\xspace} & Rank & \textbf{\qpcr} \\
\hline
1 & Philosophers are Mortal: Inferring the Truth of Unseen Facts & 1 & Philosophers are Mortal: Inferring the Truth of Unseen Facts \\
\hline
2 & Acquiring comparative commonsense knowledge from the web & 2 & NewsQA: A Machine Comprehension Dataset \\
\hline
3 & SemEval-2012 Task 7: Choice of Plausible Alternatives: An Evaluation of Commonsense Causal Reasoning & 3 & MCTest: A Challenge Dataset for the Open-Domain Machine Comprehension of Text \\
\hline
4 & Learner: a system for acquiring commonsense knowledge by analogy & 4 & Learner: a system for acquiring commonsense knowledge by analogy \\
\hline
5 & Predicting Answer Location Using Shallow Semantic Analogical Reasoning in a Factoid Question Answering System & 5 & Deep Questions without Deep Understanding \\
\hline
6 & Planning, Executing, and Evaluating the Winograd Schema Challenge & 6 & Open Mind Common Sense: Knowledge Acquisition from the General Public \\
\hline
7 & NewsQA: A Machine Comprehension Dataset & 7 & Mind Commons : An Inquisitive Approach to Learning Common Sense \\
\hline
8 & Deep Questions without Deep Understanding & 8 & SemEval-2012 Task 7: Choice of Plausible Alternatives: An Evaluation of Commonsense Causal Reasoning \\
\hline
9 & WebBrain: Joint Neural Learning of Large-Scale Commonsense Knowledge & 9 & MovieQA: Understanding Stories in Movies through Question-Answering \\
\hline
10 & A Neural Network for Factoid Question Answering over Paragraphs & 10 & Acquiring comparative commonsense knowledge from the web \\
\hline
>1000 & \textcolor{olive}{Programs with common sense} & >1000 &\textcolor{olive}{Programs with common sense} \\
\hline
\multicolumn{4}{|p{0.97\textwidth}|}{\textbf{Cited abstract:} This paper discusses programs to manipulate in a suitable formal language (most likely a part of the predicate calculus) common instrumental statements. The basic program will draw immediate conclusions from a list of premises. These conclusions will be either declarative or imperative sentences. When an imperative sentence is deduced the program takes a corresponding action. These actions may include printing sentences, moving sentences on lists, and reinitiating the basic deduction process on these lists.} \\
\hline
\end{tabular}
\caption{An example of a paragraph that cites an article that is older than the citing. Citation marker in the paragraph is printed in \textcolor{olive}{olive}, as well as the title of the cited article. Although an experienced researcher might conclude from the title and abstract alone that the cited article might be suitable for citing in recent articles on commonsense reasoning, the vocabulary mismatch between the articles might introduce difficulties for CR models.}
\label{tab:example-2.1}
\end{table*}

\begin{table*}
\small
\centering
\begin{tabular}{|p{0.05\textwidth}|p{0.38\textwidth}|p{0.05\textwidth}|p{0.38\textwidth}|}
\hline
\multicolumn{4}{|l|}{\textbf{Title:} BanditSum: Extractive Summarization as a Contextual Bandit} \\
\hline
\multicolumn{4}{|l|}{\textbf{Year:} 2018} \\
\hline
\multicolumn{4}{|p{0.97\textwidth}|}{\textbf{Abstract:} In this work, we propose a novel method for training neural networks to perform single-document extractive summarization without heuristically-generated extractive labels. We call our approach BanditSum as it treats extractive summarization as a contextual bandit (CB) problem, where the model receives a document to summarize (the context), and chooses a sequence of sentences to include in the summary (the action). A policy gradient reinforcement learning algorithm is used to train the model to select sequences of sentences that maximize ROUGE score. We perform a series of experiments demonstrating that BanditSum is able to achieve ROUGE scores that are better than or comparable to the state-of-the-art for extractive summarization, and converges using significantly fewer update steps than competing approaches. In addition, we show empirically that BanditSum performs significantly better than competing approaches when good summary sentences appear late in the source document.} \\
\hline
\multicolumn{4}{|p{0.97\textwidth}|}{\textbf{Paragraph:} \textit{Extractive summarization has been widely studied in the past.} Recently, neural network-based methods have been gaining popularity over classical methods (\textcolor{olive}{Luhn, 1958}; Gong and Liu, 2001; Conroy and O'leary, 2001; Mihalcea and Tarau, 2004; Wong et al., 2008) , as they have demonstrated stronger performance on large corpora. Central to the neural network-based models is the encoderdecoder structure. These models typically use either a convolution neural network (Kalchbrenner et al., 2014; Kim, 2014; Yin and Pei, 2015; Cao et al., 2015) , a recurrent neural network (Chung et al., 2014; Cheng and Lapata, 2016; Nallapati et al., 2017) , or a combination of the two (Narayan et al., 2018; Wu and Hu, 2018) to create sentence and document representations, using word embeddings (Mikolov et al., 2013; Pennington et al., 2014) to represent words at the input level. These vectors are then fed into a decoder network to generate the output summary. The use of reinforcement learning (RL) in extractive summarization was first explored by Ryang and Abekawa (2012) , who proposed to use the TD($\lambda$) algorithm to learn a value function for sentence selection. Rioux et al. (2014) improved this framework by replacing the learning agent with another TD($\lambda$) algorithm. However, the performance of their methods was limited by the use of shallow function approximators, which required performing a fresh round of reinforcement learning for every new document to be summarized. The more recent work of Paulus et al. (2018) and Wu and Hu (2018) use reinforcement learning in a sequential labeling setting to train abstractive and extractive summarizers, respectively, while Chen and Bansal (2018) combines both approaches, applying abstractive summarization to a set of sentences extracted by a pointer network (Vinyals et al., 2015) trained via REINFORCE. However, pre-training with a maximum likelihood objective is required in all of these models.}\\
\hline
Rank & \textbf{\textsc{SciNCL-ts}\xspace} & Rank & \textbf{\qpcr} \\
\hline
1 & Classify or Select: Neural Architectures for Extractive Document Summarization & 1 & Extractive Multi-Document Summarization with Integer Linear Programming and Support Vector Regression \\
\hline
2 & Neural Summarization by Extracting Sentences and Words & 2 & Abstractive Text Summarization Using Sequence-to-Sequence RNNs and Beyond \\
\hline
3 & A Neural Attention Model for Abstractive Sentence Summarization & 3 & Neural Summarization by Extracting Sentences and Words \\
\hline
4 & Abstractive Text Summarization Using Sequence-to-Sequence RNNs and Beyond & 4 & Multi-Document Summarization Using A* Search and Discriminative Learning \\
\hline
5 & A Neural Attention Model for Sentence Summarization & 5 & SummaRuNNer: A Recurrent Neural Network based Sequence Model for Extractive Summarization of Documents \\
\hline
6 & SummaRuNNer: A Recurrent Neural Network based Sequence Model for Extractive Summarization of Documents & 6 & Discovery of Topically Coherent Sentences for Extractive Summarization \\
\hline
7 & Framework of Automatic Text Summarization Using Reinforcement Learning & 7 & Extractive Multi-Document Summaries Should Explicitly Not Contain Document Specific Content \\
\hline
8 & Towards the Use of Deep Reinforcement Learning with Global Policy For Query-based Extractive Summarisation & 8 & Multi-document abstractive summarization using ILP based multi-sentence compression \\
\hline
9 & From neural sentence summarization to headline generation: a coarse-to-fine approach & 9 & A Hybrid Hierarchical Model for Multi-Document Summarization \\
\hline
10 & Using Supervised Bigram-based ILP for Extractive Summarization & 10 & Ultra-Summarization: A Statistical Approach to Generating Highly Condensed Non-Extractive Summaries \\
\hline
>1000 & \textcolor{olive}{The automatic creation of literature abstracts} & >1000 &\textcolor{olive}{The automatic creation of literature abstracts} \\
\hline
\multicolumn{4}{|p{0.97\textwidth}|}{\textbf{\textcolor{olive}{Cited abstract:}} Excerpts of technical papers and magazine articles that serve the purposes of conventional abstracts have been created entirely by automatic means. In the exploratory research described, the complete text of an article in machine-readable form is scanned by an IBM 704 data-processing machine and analyzed in accordance with a standard program. Statistical information derived from word frequency and distribution is used by the machine to compute a relative measure of significance, first for individual words and then for sentences. Sentences scoring highest in significance are extracted and printed out to become the "auto-abstract."} \\
\hline
\end{tabular}
\caption{Another example of an article that cites an older work. Similarly as in Table \ref{tab:example-2.1}, an experienced researcher might find the cited article fairly relevant as a classical method for extractive summarization, but CR models might struggle with vocabulary mismatch (e.g., ``model'' in citing is a ``machine'' in the cited article, ``document'' is an ``article'', and ``summary'' is an ``excerpt'').}
\label{tab:example-2.2}
\end{table*}

\begin{table*}
\small
\centering
\begin{tabular}{|p{0.05\textwidth}|p{0.38\textwidth}|p{0.05\textwidth}|p{0.38\textwidth}|}
\hline
\multicolumn{4}{|l|}{\textbf{Title:} MOROCO: The Moldavian and Romanian Dialectal Corpus} \\
\hline
\multicolumn{4}{|l|}{\textbf{Year:} 2019} \\
\hline
\multicolumn{4}{|p{0.97\textwidth}|}{\textbf{Abstract:} In this work, we introduce the MOldavian and ROmanian Dialectal COrpus (MOROCO), which is freely available for download at this https URL. The corpus contains 33564 samples of text (with over 10 million tokens) collected from the news domain. The samples belong to one of the following six topics: culture, finance, politics, science, sports and tech. The data set is divided into 21719 samples for training, 5921 samples for validation and another 5924 samples for testing. For each sample, we provide corresponding dialectal and category labels. This allows us to perform empirical studies on several classification tasks such as (i) binary discrimination of Moldavian versus Romanian text samples, (ii) intra-dialect multi-class categorization by topic and (iii) cross-dialect multi-class categorization by topic. We perform experiments using a shallow approach based on string kernels, as well as a novel deep approach based on character-level convolutional neural networks containing Squeeze-and-Excitation blocks. We also present and analyze the most discriminative features of our best performing model, before and after named entity removal.} \\
\hline
\multicolumn{4}{|p{0.97\textwidth}|}{\textbf{Paragraph:} \textit{Other languages.} The Nordic Dialect Corpus (Johannessen et al., 2009 ) contains about 466K spoken words from Denmark, Faroe Islands, Iceland, Norway and Sweden. The authors transcribed each dialect by the standard official orthography of the corresponding country. Francom et al. (2014) introduced the ACTIV-ES corpus, which represents a cross-dialectal record of the informal language use of Spanish speakers from Argentina, Mexico and Spain. The data set is composed of 430 TV or movie subtitle files. The DSL corpus collection (Tan et al., 2014) comprises news data from various corpora to emulate the diverse news content across different languages. The collection is comprised of six language vari- ety groups. For each language, the collection contains 18K training sentences, 2K validation sentences and 1K test sentences. The ArchiMob corpus (Samardžić et al., 2016) contains manuallyannotated transcripts of Swiss German speech collected from four different regions: Basel, Bern, Lucerne and Zurich. The data set was used in the 2017 and 2018 VarDial Evaluation Campaigns (Zampieri et al., 2017 (Zampieri et al., , 2018 . Kumar et al. (2018) constructed a corpus of five Indian dialects consisting of 307K sentences. The samples were collected by scanning, passing through an OCR engine and proofreading printed stories, novels and essays from books, magazines or newspapers. Romanian. To our knowledge, the only empirical study on Romanian dialect identification was conducted by Ciobanu and Dinu (2016) . In their work, Ciobanu and Dinu (2016) used only a short list of 108 parallel words in a binary classification task in order to discriminate between Daco-Romanian words versus Aromanian, Istro-Romanian and Megleno-Romanian words. Different from Ciobanu and Dinu (2016) , we conduct a large scale study on 33K documents that contain a total of about 10 million tokens.}\\
\hline
Rank & \textbf{\textsc{SciNCL-ts}\xspace} & Rank & \textbf{\qpcr} \\
\hline
1 & German Dialect Identification Using Classifier Ensembles & 1 & Exploiting Convolutional Neural Networks for Phonotactic Based Dialect Identification \\
\hline
2 & Classifying English Documents by National Dialect & 2 & German Dialect Identification Using Classifier Ensembles \\
\hline
3 & Exploiting Convolutional Neural Networks for Phonotactic Based Dialect Identification & 3 & Native Language Identification with String Kernels \\
\hline
4 & A Character-level Convolutional Neural Network for Distinguishing Similar Languages and Dialects & 4 & Discriminating between Indo-Aryan Languages Using SVM Ensembles \\
\hline
5 & CLUZH at VarDial GDI 2017: Testing a Variety of Machine Learning Tools for the Classification of Swiss German Dialects & 5 & UnibucKernel Reloaded: First Place in Arabic Dialect Identification for the Second Year in a Row \\
\hline
6 & The GW/LT3 VarDial 2016 Shared Task System for Dialects and Similar Languages Detection & 6 & Convolutional Neural Networks and Language Embeddings for End-to-End Dialect Recognition \\
\hline
7 & Sentence Level Dialect Identification in Arabic & 7 & A Character-level Convolutional Neural Network for Distinguishing Similar Languages and Dialects \\
\hline
8 & Word-Based Dialect Identification with Georeferenced Rules & 8 & Cross-corpus Native Language Identification via Statistical Embedding \\
\hline
9 & UnibucKernel Reloaded: First Place in Arabic Dialect Identification for the Second Year in a Row & 9 & An Exploration of Language Identification Techniques for the Dutch Folktale Database \\
\hline
10 & Cross-corpus Native Language Identification via Statistical Embedding & 10 & Feature Extraction for Native Language Identification Using Language Modeling \\
\hline
\end{tabular}
\caption{An example of recommendations for the paragraph describing work on dialect identification in other languages. None of the recommendations of \textsc{SciNCL-ts}\xspace and \qpcr was cited in the paragraph, although some of the articles could fit into the paragraph's narrative. }
\label{tab:example-3}
\end{table*}

\begin{table*}
\small
\centering
\begin{tabular}{|p{0.05\textwidth}|p{0.38\textwidth}|p{0.05\textwidth}|p{0.38\textwidth}|}
\hline
\multicolumn{4}{|l|}{\textbf{Title:} Introducing two Vietnamese Datasets for Evaluating Semantic Models of (Dis-)Similarity and Relatedness} \\
\hline
\multicolumn{4}{|l|}{\textbf{Year:} 2018} \\
\hline
\multicolumn{4}{|p{0.97\textwidth}|}{\textbf{Abstract:} We present two novel datasets for the low-resource language Vietnamese to assess models of semantic similarity: ViCon comprises pairs of synonyms and antonyms across word classes, thus offering data to distinguish between similarity and dissimilarity. ViSim-400 provides degrees of similarity across five semantic relations, as rated by human judges. The two datasets are verified through standard co-occurrence and neural network models, showing results comparable to the respective English datasets.} \\
\hline
\multicolumn{4}{|p{0.97\textwidth}|}{\textbf{Paragraph:} \textit{For other languages, only a few gold standard sets with scored word pairs exist.} Among others, Gurevych (2005) replicated Rubenstein and Goodenough's experiments after translating the original 65 word pairs into German. In later work, Gurevych (2006) used the same experimental setup to increase the number of word pairs to 350. Leviant and Reichart (2015) translated two prominent evaluation sets, WordSim-353 (association) and SimLex-999 (similarity) from English to Italian, German and Russian, and collected the scores for each dataset from the respective native speakers via crowdflower 6 .}\\
\hline
Rank & \textbf{\textsc{SciNCL-ts}\xspace} & Rank & \textbf{\qpcr} \\
\hline
1 & Comparing Semantic Relatedness between Word Pairs in Portuguese Using Wikipedia & 1 & Comparing Semantic Relatedness between Word Pairs in Portuguese Using Wikipedia \\
\hline
2 & Gathering Information About Word Similarity from Neighbor Sentences & 2 & A Framework for the Construction of Monolingual and Cross-lingual Word Similarity Datasets \\
\hline
3 & Duluth : Measuring Degrees of Relational Similarity with the Gloss Vector Measure of Semantic Relatedness & 3 & Human and Machine Judgements for Russian Semantic Relatedness \\
\hline
4 & A Study on Similarity and Relatedness Using Distributional and WordNet-based Approaches & 4 & SemEval-2014 Task 3: Cross-Level Semantic Similarity \\
\hline
5 & ExB Themis: Extensive Feature Extraction from Word Alignments for Semantic Textual Similarity & 5 & An English-Chinese Cross-lingual Word Semantic Similarity Measure Exploring Attributes and Relations \\
\hline
6 & Lemon and Tea Are Not Similar: Measuring Word-to-Word Similarity by Combining Different Methods & 6 & ExB Themis: Extensive Feature Extraction from Word Alignments for Semantic Textual Similarity \\
\hline
7 & Measuring Semantic Relatedness Using People and WordNet & 7 & HHU at SemEval-2016 Task 1: Multiple Approaches to Measuring Semantic Textual Similarity \\
\hline
8 & Measuring Semantic Relatedness using Multilingual Representations & 8 & A critique of word similarity as a method for evaluating distributional semantic models \\
\hline
9 & Comparing Wikipedia and German Wordnet by Evaluating Semantic Relatedness on Multiple Datasets & 9 & Gathering Information About Word Similarity from Neighbor Sentences \\
\hline
10 & A New Set of Norms for Semantic Relatedness Measures & 10 & A Study on Similarity and Relatedness Using Distributional and WordNet-based Approaches \\
\hline
\end{tabular}
\caption{An example of recommendations for the paragraph describing work in other languages on the construction of word pairs similarity datasets. None of the recommendations of \textsc{SciNCL-ts}\xspace and \qpcr were cited in the paragraph, although some of the articles could fit into the paragraph's narrative.}
\label{tab:example-4}
\end{table*}

\begin{table*}
\small
\centering
\begin{tabular}{|p{0.05\textwidth}|p{0.38\textwidth}|p{0.05\textwidth}|p{0.38\textwidth}|}
\hline
\multicolumn{4}{|l|}{\textbf{Title:} Part-of-Speech Tagging on an Endangered Language: a Parallel Griko-Italian Resource} \\
\hline
\multicolumn{4}{|l|}{\textbf{Year:} 2018} \\
\hline
\multicolumn{4}{|p{0.97\textwidth}|}{\textbf{Abstract:} Most work on part-of-speech (POS) tagging is focused on high resource languages, or examines low-resource and active learning settings through simulated studies. We evaluate POS tagging techniques on an actual endangered language, Griko. We present a resource that contains 114 narratives in Griko, along with sentence-level translations in Italian, and provides gold annotations for the test set. Based on a previously collected small corpus, we investigate several traditional methods, as well as methods that take advantage of monolingual data or project cross-lingual POS tags. We show that the combination of a semi-supervised method with cross-lingual transfer is more appropriate for this extremely challenging setting, with the best tagger achieving an accuracy of 72.9\%. With an applied active learning scheme, which we use to collect sentence-level annotations over the test set, we achieve improvements of more than 21 percentage points.} \\
\hline
\multicolumn{4}{|p{0.97\textwidth}|}{\textbf{Paragraph:} \textit{Most of the previous approaches are rarely tested on under-represented languages, with research on POS tagging for endangered languages being sporadic.} Ptaszynski and Momouchi (2012) , for example, presented an HMM-based POS tagger for the extremely endangered Ainu language, based on dictionaries, 12 narratives (yukar), using one annotated story (200 words) for evaluation. To our knowledge, no other previous work has extensively tested several approaches on an actual endangered language.}\\
\hline
Rank & \textbf{\textsc{SciNCL-ts}\xspace} & Rank & \textbf{\qpcr} \\
\hline
1 & CombiTagger: A System for Developing Combined Taggers & 1 & Distantly Supervised POS Tagging of Low-Resource Languages under Extreme Data Sparsity: The Case of Hittite \\
\hline
2 & Distantly Supervised POS Tagging of Low-Resource Languages under Extreme Data Sparsity: The Case of Hittite & 2 & What Can We Get From 1000 Tokens? A Case Study of Multilingual POS Tagging For Resource-Poor Languages \\
\hline
3 & TagMiner: A Semisupervised Associative POS Tagger Effective for Resource Poor Languages & 3 & Unsupervised adaptation of supervised part-of-speech taggers for closely related languages \\
\hline
4 & Improving the PoS tagging accuracy of Icelandic text & 4 & TagMiner: A Semisupervised Associative POS Tagger Effective for Resource Poor Languages \\
\hline
5 & Wiki-ly Supervised Part-of-Speech Tagging & 5 & Part-of-Speech Tag Disambiguation by Cross-Linguistic Majority Vote \\
\hline
6 & Coupling an annotated corpus and a lexicon for state-of-the-art POS tagging & 6 & POS Tagging for Historical Texts with Sparse Training Data \\
\hline
7 & Something Borrowed, Something Blue: Rule-based Combination of POS Taggers & 7 & Unsupervised and Lightly Supervised Part-of-Speech Tagging Using Recurrent Neural Networks \\
\hline
8 & Coupling an annotated corpus and a morphosyntactic lexicon for state-ofthe-art POS tagging with less human effort & 8 & Wiki-ly Supervised Part-of-Speech Tagging \\
\hline
9 & What Can We Get From 1000 Tokens? A Case Study of Multilingual POS Tagging For Resource-Poor Languages & 9 & Improving Accuracy in Word Class Tagging through the Combination of Machine Learning Systems \\
\hline
10 & Ensemble system for Part-of-Speech tagging & 10 & Mac-Morpho Revisited: Towards Robust Part-of-Speech Tagging \\
\hline
\end{tabular}
\caption{An example of recommendations for the paragraph describing work on POS tagging for endangered languages. Similar to Table \ref{tab:example-4}, none of the recommendations of \textsc{SciNCL-ts}\xspace and \qpcr were cited in the paragraph, although some of the articles could fit into the paragraph's narrative.}
\label{tab:example-5}
\end{table*}

\begin{table*}
\small
\centering
\begin{tabular}{|p{0.05\textwidth}|p{0.38\textwidth}|p{0.05\textwidth}|p{0.38\textwidth}|}
\hline
\multicolumn{4}{|l|}{\textbf{Title:} Dependency Parsing for Spoken Dialog Systems} \\
\hline
\multicolumn{4}{|l|}{\textbf{Year:} 2019} \\
\hline
\multicolumn{4}{|p{0.97\textwidth}|}{\textbf{Abstract:} Dependency parsing of conversational input can play an important role in language understanding for dialog systems by identifying the relationships between entities extracted from user utterances. Additionally, effective dependency parsing can elucidate differences in language structure and usage for discourse analysis of human-human versus human-machine dialogs. However, models trained on datasets based on news articles and web data do not perform well on spoken human-machine dialog, and currently available annotation schemes do not adapt well to dialog data. Therefore, we propose the Spoken Conversation Universal Dependencies (SCUD) annotation scheme that extends the Universal Dependencies (UD) (Nivre et al., 2016) guidelines to spoken human-machine dialogs. We also provide ConvBank, a conversation dataset between humans and an open-domain conversational dialog system with SCUD annotation. Finally, to demonstrate the utility of the dataset, we train a dependency parser on the ConvBank dataset. We demonstrate that by pre-training a dependency parser on a set of larger public datasets and fine-tuning on ConvBank data, we achieved the best result, 85.05\% unlabeled and 77.82\% labeled attachment accuracy.} \\
\hline
\multicolumn{4}{|p{0.97\textwidth}|}{\textbf{Paragraph:} \textit{Previous work has demonstrated that expanding the UD annotation scheme can result in successful parsers for other domains.} For example, Liu et al. (2018) expand the UD scheme to train a parser for Twitter data. Thus we take a similar approach in expanding the UD scheme to encompass issues common to automated speech transcription.}\\
\hline
Rank & \textbf{\textsc{SciNCL-ts}\xspace} & Rank & \textbf{\qpcr} \\
\hline
1 & Adding Syntactic Annotations to Transcripts of Parent-Child Dialogs & 1 & Discourse parsing for multi-party chat dialogues \\
\hline
2 & Discourse parsing for multi-party chat dialogues & 2 & Annotation for and Robust Parsing of Discourse Structure on Unrestricted Texts \\
\hline
3 & A Dependency Treebank for English & 3 & A Dependency Parser for Tweets \\
\hline
4 & Discovering Conversational Dependencies between Messages in Dialogs & 4 & An Iterative Approach for Joint Dependency Parsing and Semantic Role Labeling \\
\hline
5 & Annotating Spoken Dialogs: From Speech Segments to Dialog Acts and Frame Semantics & 5 & Multilingual Dependency Learning: A Huge Feature Engineering Method to Semantic Dependency Parsing \\
\hline
6 & DailyDialog: A Manually Labelled Multi-turn Dialogue Dataset & 6 & SemEval-2012 Task 5: Chinese Semantic Dependency Parsing \\
\hline
7 & The ICSI Meeting Recorder Dialog Act (MRDA) Corpus & 7 & Parsing Tweets into Universal Dependencies \\
\hline
8 & The Tübingen Treebanks for Spoken German, English, and Japanese & 8 & Hybrid Learning of Dependency Structures from Heterogeneous Linguistic Resources \\
\hline
9 & Annotation for and Robust Parsing of Discourse Structure on Unrestricted Texts & 9 & Universal Dependency Parsing with a General Transition-Based \\
\hline
10 & Dialogue Act Modeling for Automatic Tagging and Recognition of Conversational Speech & 10 & Dependency Parsing: Past, Present, and Future \\
\hline
\end{tabular}
\caption{An example of recommendations demonstrating high topical relevancy of recommended titles for the topic sentence in the case of \qpcr. While \textsc{SciNCL-ts}\xspace's recommendations are more focused on dialog systems, as that is the main topic of the citing article, \qpcr's recommendations are more about universal dependencies, which are mentioned in the topic sentence.}
\label{tab:example-6}
\end{table*}

\begin{table*}
\small
\centering
\begin{tabular}{|p{0.05\textwidth}|p{0.38\textwidth}|p{0.05\textwidth}|p{0.38\textwidth}|}
\hline
\multicolumn{4}{|l|}{\textbf{Title:} Cross-lingual Argumentation Mining: Machine Translation (and a bit of Projection) is All You Need!} \\
\hline
\multicolumn{4}{|l|}{\textbf{Year:} 2018} \\
\hline
\multicolumn{4}{|p{0.97\textwidth}|}{\textbf{Abstract:} Argumentation mining (AM) requires the identification of complex discourse structures and has lately been applied with success monolingually. In this work, we show that the existing resources are, however, not adequate for assessing cross-lingual AM, due to their heterogeneity or lack of complexity. We therefore create suitable parallel corpora by (human and machine) translating a popular AM dataset consisting of persuasive student essays into German, French, Spanish, and Chinese. We then compare (i) annotation projection and (ii) bilingual word embeddings based direct transfer strategies for cross-lingual AM, finding that the former performs considerably better and almost eliminates the loss from cross-lingual transfer. Moreover, we find that annotation projection works equally well when using either costly human or cheap machine translations. Our code and data are available at http://github.com/UKPLab/coling2018-xling\_argument\_mining.} \\
\hline
\multicolumn{4}{|p{0.97\textwidth}|}{\textbf{Paragraph:} \textit{Cross-lingual Word Embeddings are the (modern) basis of the direct transfer method.} As with monolingual embeddings, there exists a veritable zoo of different approaches, but they often perform very similarly in applications (Upadhyay et al., 2016) and seemingly very different approaches are oftentimes also equivalent on a theoretical level (Ruder et al., 2017) .}\\
\hline
Rank & \textbf{\textsc{SciNCL-ts}\xspace} & Rank & \textbf{\qpcr} \\
\hline
1 & A Two-Phase Approach Towards Identifying Argument Structure in Natural Language & 1 & Inducing Multilingual Text Analysis Tools Using Bidirectional Recurrent Neural Networks \\
\hline
2 & Argumentative Writing Support by means of Natural Language Processing & 2 & Cross-Lingual Induction and Transfer of Verb Classes Based on Word Vector Space Specialisation \\
\hline
3 & Baselines and test data for cross-lingual inference & 3 & Cross-lingual sentiment transfer with limited resources \\
\hline
4 & Using Discourse Structure Improves Machine Translation Evaluation & 4 & Transferring Coreference Resolvers with Posterior Regularization \\
\hline
5 & A Shared Task on Argumentation Mining in Newspaper Editorials & 5 & Leveraging Monolingual Data for Crosslingual Compositional Word Representations \\
\hline
6 & Neural End-to-End Learning for Computational Argumentation Mining & 6 & Learning Crosslingual Word Embeddings without Bilingual Corpora \\
\hline
7 & Exploiting Debate Portals for Semi-Supervised Argumentation Mining in User-Generated Web Discourse & 7 & One-Shot Neural Cross-Lingual Transfer for Paradigm Completion \\
\hline
8 & Discourse Structure in Machine Translation Evaluation & 8 & Evaluating Unsupervised Dutch Word Embeddings as a Linguistic Resource \\
\hline
9 & The Far Reach of Multiword Expressions in Educational Technology & 9 & Bilingual Word Embeddings from Parallel and Non-parallel Corpora for Cross-Language Text Classification \\
\hline
10 & Discourse-level features for statistical machine translation & 10 & A Strong Baseline for Learning Cross-Lingual Word Embeddings from Sentence Alignments \\
\hline
\end{tabular}
\caption{Another example of recommendations that demonstrate high topical relevancy of recommended titles for the topic sentence in the case of \qpcr. \textsc{SciNCL-ts}\xspace's recommendations dominantly cover topics of argumentation mining and machine translation, while \qpcr's recommendations are about the cross-lingual transfer, i.e., focused on the topic of the topic sentence.}
\label{tab:example-7}
\end{table*}

\subsection{Model Training Details}

\scincl, the model that we compare with and fine-tune, is a transformer-based model whose architecture resembles BERT \citep{devlin-etal-2019-bert}.
The model we used contains 12 layers with hidden state embeddings of size 768.
We fine-tuned all the models on an NVIDIA GeForce RTX 3090 GPU with 24GB of RAM. 
We used PyTorch version 1.12.1, Transformers version 4.20.1, CUDA 11.4, and Faiss 1.7.2.
In total, all experiments and evaluations amounted to around 125 GPU hours.

\end{document}